\documentclass[12pt]{spieman}  
\pdfoutput=1
\usepackage{amsmath,amsfonts,amssymb}
\usepackage{graphicx}
\usepackage{setspace}
\usepackage{tocloft}
\usepackage{caption} 

\title{A Value Mapping Virtual Staining Framework for Large-scale Histological Imaging}

\author[a,e]{Junjia Wang}
\author[b,c,e]{Bo Xiong}
\author[d,*]{You Zhou}
\author[a]{Xun Cao}
\author[a]{Zhan Ma}
\affil[a]{School of Electronic Science and Engineering, Nanjing University, Nanjing, 210023, China.}
\affil[b]{National Engineering Research Center of Visual Technology, Peking University, Beijing, 100871, China.}
\affil[c]{School of Computer Science, Peking University, Beijing, 100871, China.}
\affil[d]{Medical School, Nanjing University, Nanjing, 210093, China.}
\affil[e]{These authors contribute equally to this work.}

\cftpagenumbersoff{figure}
\cftpagenumbersoff{table} 
\begin{document} 
\maketitle

\begin{abstract}
The emergence of virtual staining technology provides a rapid and efficient alternative for researchers in tissue pathology. It enables the utilization of unlabeled microscopic samples to generate virtual replicas of chemically stained histological slices, or facilitate the transformation of one staining type into another. The remarkable performance of generative networks, such as CycleGAN, offers an unsupervised learning approach for virtual coloring, overcoming the limitations of high-quality paired data required in supervised learning. Nevertheless, large-scale color transformation necessitates processing large field-of-view images in patches, often resulting in significant boundary inconsistency and artifacts. Additionally, the transformation between different colorized modalities typically needs further efforts to modify loss functions and tune hyperparameters for independent training of networks. In this study, we introduce a general virtual staining framework that is adaptable to various conditions. We propose a loss function based on the value mapping constraint to ensure the accuracy of virtual coloring between different pathological modalities, termed the Value Mapping Generative Adversarial Network (VM-GAN). Meanwhile, we present a confidence-based tiling method to address the challenge of boundary inconsistency arising from patch-wise processing. Experimental results on diverse data with varying staining protocols demonstrate that our method achieves superior quantitative indicators and improved visual perception.
\end{abstract}


{\noindent \footnotesize\textbf{*}You Zhou, \linkable{zhouyou@nju.edu.cn} }

\begin{spacing}{2}   

\section{Introduction}
\label{sect:intro}  
Histological staining serves as a fundamental technique utilized in disease diagnosis and pathology research\cite{Bancroft2008}. By employing specific markers tailored to label different biological elements according to their inherent characteristics, histological staining enables the visualization of distinct tissue and cellular structures, allowing for subsequent pathological analysis and disease diagnosis. 
The application of diverse staining techniques highlights a range of biological features. Hematoxylin and eosin (H$\&$E) staining, a prominent method on histopathology slides, provides rich information and is widely used in tumor diagnosis\cite{Titford2005,Coudray2018}. Immunohistochemistry (IHC) staining represents a technique that integrates immunology with traditional histology, enabling the detection and localization of specific chemical substances within tissues and cells through the specific binding interaction between antigens and antibodies\cite{Ramos-Vara2014}. For instance, Ki67 IHC staining, commonly used to evaluate cell proliferation activity, aids pathologists and researchers in quantitatively assessing the extent of tumor proliferation. However, these standard histological staining procedures are typically executed in pathological laboratories, involving multiple sample preparation workflows that are both labor-intensive and time-consuming.

Recently, deep learning has made remarkable advancements in the fields of light microscopy\cite{Barbastathis2019,Weigert2018,Wang2019,Ouyang2018,Wu2019,Ounkomol2018}, including applications in virtual histological staining\cite{Christiansen2018,Zhu2017,Bai2023,Boyd2022,Liu2021,Wang2020Virtual,Li2021}. Virtual staining circumvents the need for laborious sample preparation, specialized laboratory equipment, and technical personnel requirement in traditional chemical histological staining\cite{Bancroft2008,Musumeci2014}. By leveraging deep learning for digital histological staining generation, virtual staining offers an efficient, precise, and cost-effective alternative.
Deep learning-based virtual staining methods are primarily categorized into supervised and unsupervised approaches. Supervised methods\cite{Christiansen2018} rely on a large amount of high-quality paired images; however, providing complete one-to-one correspondences between two data modalities is often challenging in light microscopic applications. Hence, unsupervised learning methods based on cycle generative adversarial network (CycleGAN) \cite{Zhu2017,Bai2023,Boyd2022,Liu2021,Wang2020Virtual,Li2021} are highly favored by relevant researchers due to their practical value in clinical imaging and scientific research. 
For instance, the multi-scale Structural Similarity (SSIM) loss is applied to constrain the green channel between different color modalities, ensuring structural consistency after modal conversion \cite{Wang2020Virtual}. The Unsupervised content-preserving Transformation for Optical Microscopy (UTOM) method is proposed to maintain the organizational structure transformation accuracy through a mask extracting step\cite{Li2021}. But when faced with the re-staining task, UTOM is prone to distortion. Liu et al. propose a pathology consistency constraint based on CycleGAN and combines it with structural similarity constraints, achieving a high-quality stain transfer between unpaired H$\&$E and Ki67 stained images \cite{Liu2021}. Lee et al. combine the physical parameterized model with a generative network and utilize the concept of cycle-consistency loss, achieving accurate transformation from diffraction patterns to holographic imaging in complex and dynamic environments \cite{Lee2023}. Stimulated Raman CycleGAN (SRC-GAN) \cite{Liu2024} designs a semi-supervised algorithm that adapts the network architecture of CycleGAN specifically for virtual staining of Stimulated Raman Scattering (SRS) images, which however requires paired datasets as the supervised component.

While several existing works have achieved good performance in virtual staining, they also encounter challenges when processing large-scale and high-resolution images, which are often necessary in microscopic applications. Additionally, color transformation between different histological modalities generally requires different loss functions and parameters in network design, limiting its widespread application. Specifically, for the practical use of unsupervised deep learning networks in microscopic image modal transformation, the following challenges need to be addressed: (1) The identity loss in CycleGAN is unstable in microscopic scenes, potentially leading to color reversal during staining. (2) Large-scale and high-resolution images must be cropped into patches for network training and inference, introducing inconsistencies and artifacts in color, brightness, and contrast between these patches, especially at the edges of each patch. (3) Specific cross-modality transformations require corresponding network designs, which currently lack a relatively general framework. 

To address these issues, in this paper, we present a value mapping virtual staining framework based on the CycleGAN network, abbreviated as VM-GAN, for histological imaging. We propose to convert the Red-Green-Blue (RGB) color space of images to Hue-Saturation-Value (HSV) color space and then extract the value channel as a novel constraint, ensuring correct color mapping relationships and preservation of image content during modality transformation. We also adopt a confidence-based splicing post-processing method to eliminate artifacts generated in patch splicing in a plug-and-play manner. This framework is capable of handling various cross-modality transformations and is well-suited for processing large-scale and high-resolution images, addressing the boundary inconsistency issues encountered in many existing methods. We conduct several experiments using diverse large field-of-view (FOV) datasets with varying staining modalities, demonstrating that our method achieves superior quantitative indicators and improves visual perception compared to several commonly used methods.

\begin{figure*}[htb]
\centering
{\includegraphics[width=\linewidth]{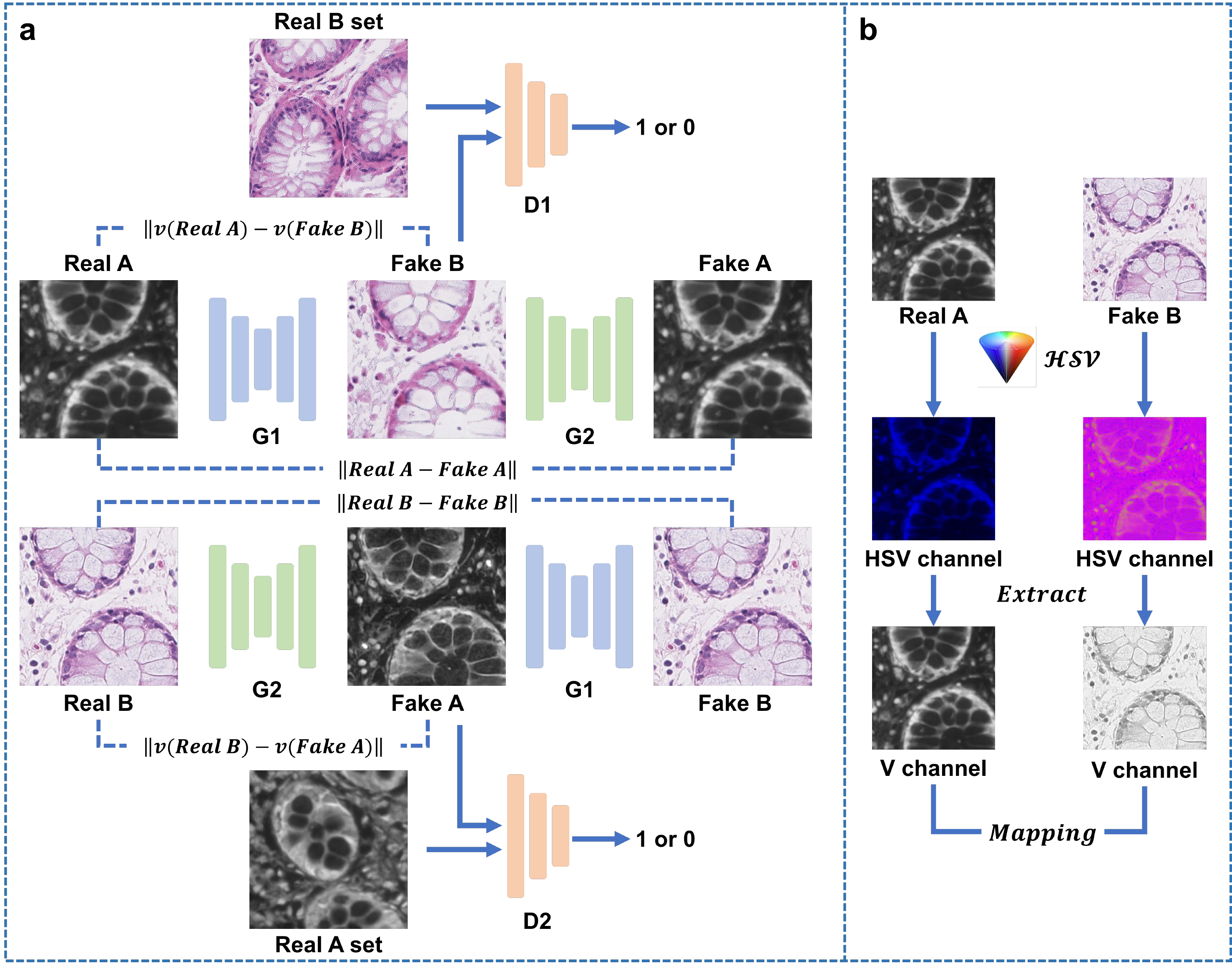}}
\caption{Architecture of the proposed VM-GAN model. (a) The main structure of the model, where the total loss function includes the adversarial loss, the cycle consistency loss, and the value loss. $\mathit{v}$(·) represents the value channel of the image. G and D correspond to the generator and discriminator respectively. (b) The process of converting from RGB color space to HSV space and extracting the value channel.}
\label{fig1}
\rule{\textwidth}{1pt}
\end{figure*}

\section{METHODS}

In our proposed method, we first transition the virtual staining process from the commonly used RGB space to the HSV space. Through the design and calculation of value loss, we can guarantee the consistency and suitability of transformation among different color modalities. We next build the general value mapping framework for large-scale virtual staining. Our method comprises the following key components: (1) HSV space conversion, (2) loss function design of the VM-GAN network, and (3) confidence tiling scheme for large-scale samples.

\subsection{HSV Space Conversion}
\label{sect:title}
The HSV color space describes how people experience color better than RGB dose \cite{Stockman2001}. It divides the attributes of color into three main components: hue, saturation, and value. Hue represents the fundamental property of color and is typically represented by an angle value ranging from $0^{\circ}$ to $360^{\circ}$, corresponding to colors in the color spectrum. For example, red is located at $0^{\circ}$ or $360^{\circ}$, green at $120^{\circ}$, and blue at $240^{\circ}$. Saturation refers to the purity or intensity of a color. Higher saturation indicates a purer color, while lower saturation indicates a color that is closer to gray. Value represents the brightness or darkness of a color, with higher values indicating brighter colors and lower values indicating darker colors. The HSV color space is fundamentally different from the well-known RGB color space because it separates brightness (value) from color information (hue), which aligns more with human perception and understanding of colors. It provides an intuitive way to describe and manipulate colors, such as adjusting brightness, saturation, and hue in computer vision fields\cite{Bargshady2020}.

The conversion formula from RGB to HSV space is as follows:
\begin{equation}
   \mathcal{V} = Max
\end{equation}
\begin{equation}
   \mathcal{S} = (Max-Min)/Max
\end{equation}
\begin{equation}
\mathcal{H}= 
\begin{cases}
\frac{G-B}{Max-Min}\times 60^{\circ}, & \text{If $R$ is $Max$} \\
\frac{B-R}{Max-Min}\times 60^{\circ}+120^{\circ}, & \text{If $G$ is $Max$} \\ 
\frac{R-G}{Max-Min}\times 60^{\circ}+240^{\circ}, & \text{If $B$ is $Max$}
\end{cases}
\end{equation}
where $V$, $S$, and $H$ represent the value, saturation, and hue components of HSV space for a given image pixel. $Max$ and $Min$ respectively indicate the maximum and minimum values in RGB channels. $R$, $G$, and $B$ refer to the pixel values of corresponding red, green, and blue channels.

\subsection{Value Mapping Virtual Staining Framework}
Based on the HSV space conversion, we propose a value mapping virtual staining method built on the CycleGAN network. We design a novel constraint by exacting the value channel of the HSV space and modify the loss function of CycleGAN accordingly. The main structure of our framework is shown in Fig.~\ref{fig1}, where we replace the identity loss in the original network with value loss to ensure the correct color mapping relationship between Domain A and Domain B. Notably, aside from calculating the value loss, the main process of our method is conducted in RGB channels separately.

According to the original CycleGAN network, we divide the unpaired images into Domain A and Domain B, thus enabling the network to undergo unsupervised training. The entire architecture includes two generators ($G_1$ and $G_2$) and two discriminators ($D_1$ and $D_2$). Generator $G_1$ takes real images from Domain A as input and generates fake B, while discriminator $D_1$ distinguishes between real images from Domain B and fake B. Each generator has a corresponding discriminator, attempting to differentiate its generated images from real images. Similarly, generator $G_2$ generates fake A based on real images from Domain B, and discriminator $D_2$ is responsible for distinguishing between real images from Domain A and fake A. The loss function of our proposed general virtual staining framework can be written as:  
\begin{equation}
\begin{aligned}
\mathcal{L}_{VM-GAN} 
& =\mathcal{L}_{GAN}(G_1, D_1, A, B) \\
& +\mathcal{L}_{GAN}(G_2, D_2, A, B) \\
& +\lambda_{\text{cycle}} \mathcal{L}_{\text{cycle}} \\
& +\lambda_{\text{value}} \| \text {value}\left(G_1(A)\right)-\operatorname{value}(A)\|_1 \\
& +\lambda_{\text{value}}\left\|\operatorname{value}\left(G_2(B)\right)-\operatorname{value}(B)\right\|_1
\end{aligned}
\label{eq_VM-GAN}
\end{equation}
where $\mathcal{L}_{GAN}(G_1, D_1, A, B)$ and $\mathcal{L}_{GAN}(G_2, D_2, A, B)$ represent the adversarial losses of $G_1$ and $G_2$ respectively, $\mathcal{L}_{\text{cycle}}$ is the cycle consistency loss, and $\lambda_{\text{cycle}}$ is the weight of the cycle consistency loss. Specifically, the adversarial loss is used to train models by minimizing the adversarial loss function between the generator and the discriminator. Through the optimization of the adversarial loss, the generator and the discriminator engage in iterative adversarial training, ultimately enabling the generator to generate images that are similar to the real images in the target domain, while the discriminator can accurately distinguish between the generated images and the real images. The cycle consistency loss is introduced to preserve the original content of an image and prevent it from being destroyed during the generation process, and is achieved by mutually reconstructing the images. By optimizing the cycle consistency loss, the generators can learn the mapping relationship between Modality A and Modality B, ensuring image consistency and integrity during modality translation. The adversarial loss and cycle consistency loss between domain A and B can be expressed as:
\begin{equation}
\begin{aligned}
\mathcal{L}_{G A N}(G_1, D_1, A, B)
&=E_{b \sim p(b)}\left[\log D_1(b)\right]\\
&+E_{a \sim p(a)}\left[\log \left(1-D_1(G_1(a))\right)\right]
\end{aligned}
\end{equation}
\begin{equation}
\begin{aligned}
\mathcal{L}_{G A N}(G_2, D_2, A, B)
&=E_{a \sim p(a)}\left[\log D_1(a)\right]\\
&+E_{b \sim p(b)}\left[\log \left(1-D_2(G_2(b))\right)\right]
\end{aligned}
\end{equation}
\begin{equation}
\begin{aligned}
\mathcal{L}_{\text {cycle }}
&=E_A\left[\left\|G_2\left(G_1(A)\right)-A\right\|_1\right]\\
&+E_B\left[\left\|G_1\left(G_2(B)\right)-B\right\|_1\right]
\end{aligned}
\end{equation}
where $E(\cdot)$ and $p(\cdot)$ represent the expectation and distribution probability of data respectively.

The last two items in Equation.~\eqref{eq_VM-GAN} are our designed value losses, where $G_1(A)$ and $G_2(B)$ correspond to the output of real A input into generator $G_1$ and real B input into generator $G_2$ respectively. Value ($\cdot$) represents the value of the value channel, $\Vert \cdot \Vert_{1}$ is the $\ell_1$ norm, and $\lambda_{\text {value}}$ is the weight of the value loss. 
We design the value loss to address the inadequate performance and weak constraints of CycleGAN in dealing with complex texture domain transformation. By optimizing both the generator $G_1$ and generator $G_2$ in the value channel of HSV space, the value loss can preserve structural integrity and avoid color reversal during the virtual staining process. Taking generator A as an example, we input a domain A image and output a fake domain B image. We convert the RGB channels of the original image into the HSV channels and extract the value channel for loss calculation. Due to its ability to avoid color reversal during color transformation, our newly-designed loss function is well suited to a wide range of pathological modalities and can serve as a general framework, without the requirement of further specialized loss function design and parameter setting.
\begin{figure*}[t]
\centering
{\includegraphics[width=0.9\linewidth]{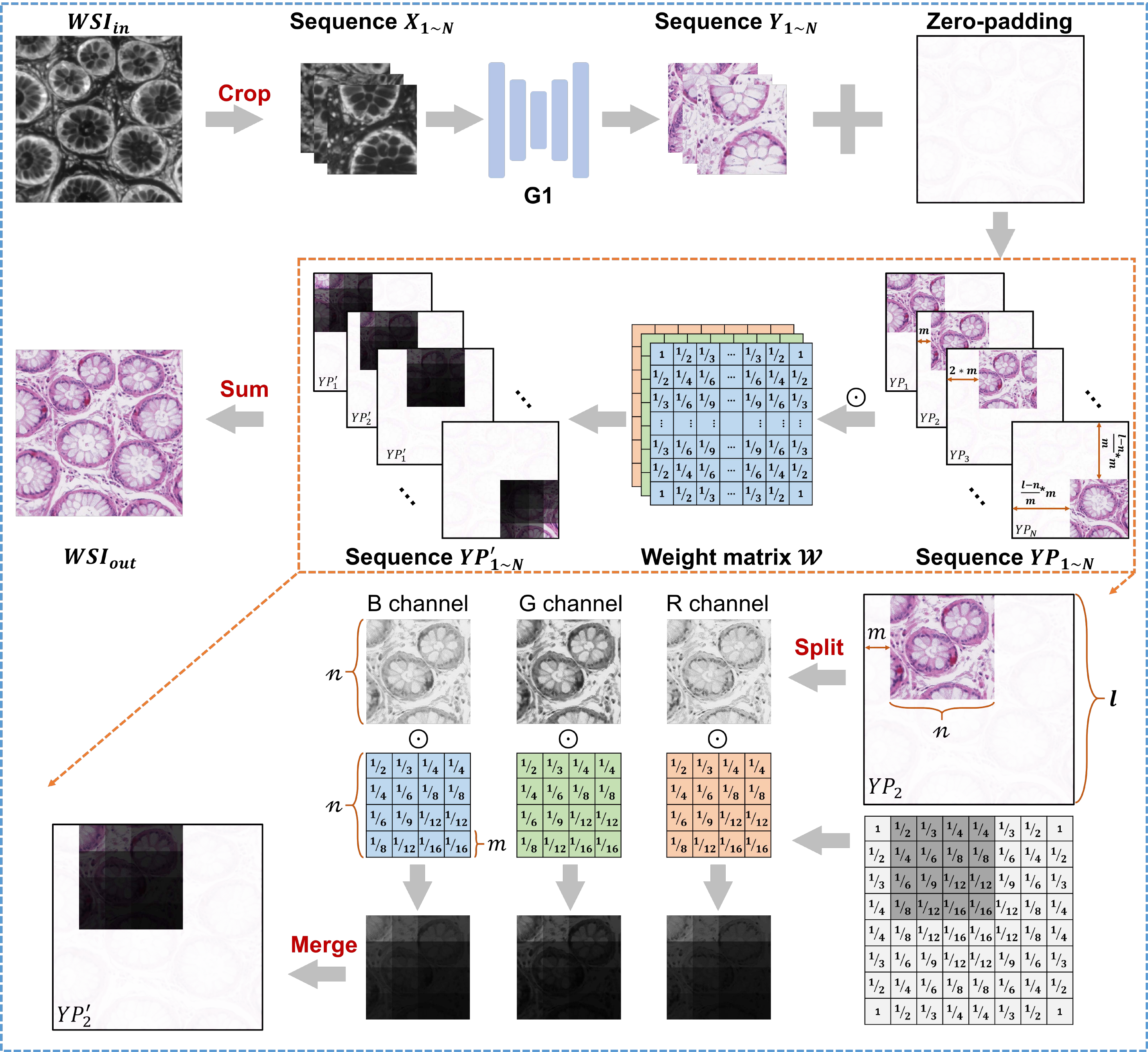}}
\caption{Confidence tiling scheme for large-scale samples. During the zero-padding process in the bottom right corner, the filling interval for each small patches is $m$ pixel. $n$ represents the size of small patches and $l$ represents the size of the $WSI_{in}$. After processing with the weight matrix, the small patches are summed up to obtain the $WSI_{out}$. The example in the lower half of the figure demonstrates in detail the process of how the weight matrix functions.}
\label{fig2}
\rule{\textwidth}{1pt}
\end{figure*}

\subsection{Confidence Tiling Scheme for Large-scale Imaging}
Regarding large-scale virtual staining, conventional methods tend to crop large images into patches for inference and directly concatenate them, leading to staining errors and structural damages due to incomplete edge structures in small images. To address this issue, we propose a confidence tiling scheme for large-scale imaging through a patch-to-whole synthesis step utilizing a weight matrix.

As shown in Fig.~\ref{fig2}, we first crop the large-scale and high-resolution whole slide image (WSI) of domain A, marked as $WSI_{in}$, into a small image Sequence $X_{1 \sim N}$ by every $m$ pixels. We input the Sequence $X_{1 \sim N}$ into the trained network model, obtaining the output Sequence $Y_{1 \sim N}$ in domain B. We then fill each image from Sequence $Y_{1 \sim N}$ into a zero-padding matrix according to its original position in $WSI_{in}$, forming the Sequence $YP_{1 \sim N}$. Next, we design a weight matrix $\mathcal{W}$ as the same size $l\times l$ as the $WSI_{in}$ based on the characteristics of patch-wise virtual staining. Finally, we multiply $\mathcal{W}$ by the Sequence $YP_{1 \sim N}$ using the Hadamard product (element-wise product), and normalize it to obtain the final output fake WSI in domain B, marked as $WSI_{out}$. The RGB channels of the large-scale image are processed independently. The overall function of the confidence tiling scheme can be expressed as:
\begin{equation}
WSI_{\text{out}}=\sum_{i=1}^N \text{Sequence}_{{YP}_i}\odot\mathcal{W}
\label{synthesized matrix}
\end{equation}
where $WSI_{\text{out}}$ corresponds to the output tiled whole slice, $\text{Sequence}_{YP_i}$ is the small patch sequence after virtual staining and zero-pading filling, $N$ is the number of small patches, and $\odot$ represents the Hadamard product. 

The weight matrix $\mathcal{W}$ is centrosymmetric and identical across the three RGB channels, which can be calculated by the following formula:
\begin{equation}
\frac{1}{\mathcal{V}_{xy}}= \begin{cases}\lceil \frac{x}{m}+1\rceil  \times\lceil \frac{y}{m}+1\rceil, & x, y<n \\ \lceil\frac{x}{m}+1\rceil\times \frac{n}{m}, & x<n, y \geq n \\  \frac{n}{m} \times \lceil\frac{y}{m}+1\rceil, & x \geq n, y<n \\ \frac{n}{m}\times \frac{n}{m}, & x \geq n, y \geq n\end{cases}
\label{weight matrix}
\end{equation}
where $\mathcal{V}_{x y}$ corresponds to the values in the x-th row and y-th column of the upper-left portion of the weight matrix $\mathcal{W}$ in all three RGB channels, $\lceil \cdot \rceil$ is the ceiling function, $n$ is the size of cropped small patches, and $m$ is the filling interval of neighbouring patches. The settings for $m$ and $n$ should ensure that $n/m$ is an integer. The design of weight matrix $\mathcal{W}$ is based on two phenomena we observed after virtual staining on small patches: (1) The edges of the small patches are more prone to distortion compared to the center, often resulting in cellular structure damage and staining errors. In other words, virtual staining in the central region is more reliable and realistic than in the edge areas. (2) There are inconsistencies in contrast between the small patches, leading to a decrease in visual quality when tiling the patches together to form the WSI. Through this post-processing method based on confidence principles, we can successfully eliminate the artifacts associated with WSIs \cite{Farahani2015} and the square effect.

\section{Experiments and results}
\label{sect:sections}
In experiments, we demonstrate the general virtual staining ability of our method through colorized transformation between three different modalities, e.g. autofluorescence to H$\&$E staining, H$\&$E to IHC(Ki67) staining, and IHC(Cc10) to IHC(CD31) staining, utilizing three public datasets \cite{Li2021,Liu2020,Borovec2020}. Autofluorescence images and their consecutive registered H$\&$E stained data are obtained from the UTOM dataset \cite{Li2021}, which includes whole-slide pathology images of various tumors. Dataset from Automatic Non-rigid Histological Image Registration (ANHIR) challenge \cite{Borovec2020} are collected for pathological image registration and virtual staining tasks, encompassing a range of tissues such as kidney, breast, lung, and stomach. For our experiments, we select lung lesion images from ANHIR dataset. We compare our method with several commonly-used unsupervised learning methods for virtual staining, including the orginal CycleGAN \cite{Zhu2017}, the method proposed by Wang et al. \cite{Wang2020Virtual}, and the UTOM method \cite{Li2021}. Depending on the computational power-performance trade-off, we set the patch size to $n=512$ and interval size to $m=128$. Throughout all experiments, we do not modify the hyperparameters of the networks, underscoring the generality of our methods.

\begin{figure*}[htbp]
\centering
{\includegraphics[width=0.95\linewidth]{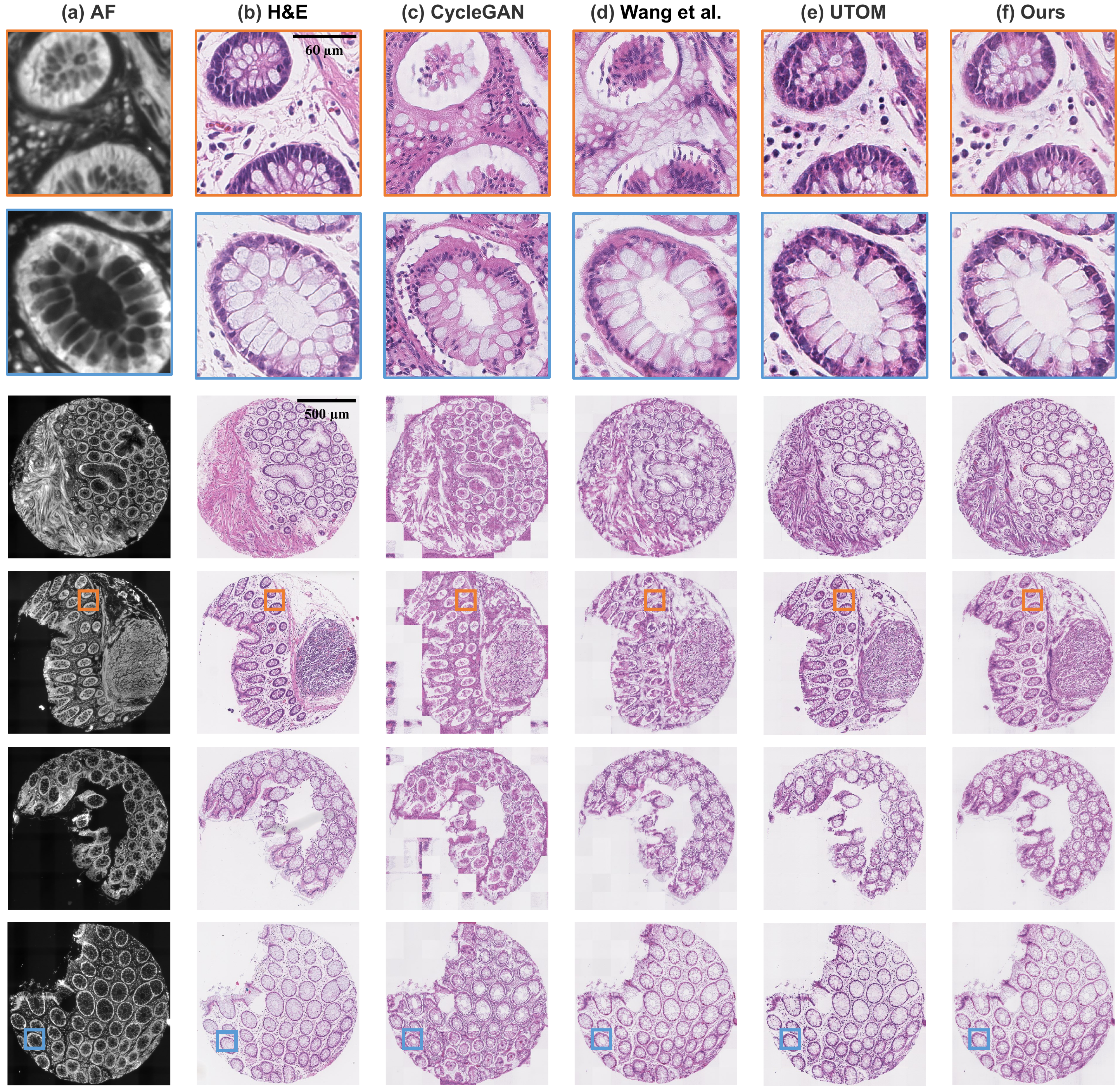}}
\caption{H$\&$E virtual staining results on autofluorescence images. (a) and (b) display the autofluorescence slides and their consecutive registered H$\&$E stained slices. (c), (d), and (e) present the H$\&$E virtual staining results obtained by the original CycleGAN \cite{Zhu2017}, the method proposed by Wang et al. \cite{Wang2020Virtual}, and the UTOM method \cite{Li2021}. (f) exhibits the H$\&$E virtual staining results obtained using our method. Enlarged images are provided in the first two rows with a scale bar of 60 $\mu$m, and whole slice images (WSIs) of different samples are shown in the last four rows with a scale bar of 500 $\mu$m. Our method effectively avoids the color reversal phenomenon in virtual staining, achieving performance comparable to that of UTOM. AF: autofluorescence.}
\label{fig_res1}
\rule{\textwidth}{1pt}
\end{figure*}

\subsection{Autofluorescence to H$\&$E}
As mentioned, in clinical practice, H$\&$E stained histopathological slides provide rich information and are widely used for tumor diagnosis\cite{Coudray2018}. Traditional histopathological imaging poses challenges for preoperative diagnosis and efficient tumor screening due to its cumbersome procedures. Researchers have long sought methods to generate standard H$\&$E stained images through virtual staining from chemical unlabeled samples \cite{Lu2016,Orringer2017,Hollon2020}.  Given that autofluorescence images generally exhibit rich histological features, converting autofluorescence to H$\&$E stained images has become increasingly relevant, aiding clinical workers in rapid diagnosis. We first demonstrate that our method can well handle the virtual staining task from autofluorescence images to their H$\&$E pairs, achieving performance comparable to the UTOM method.

The experimental results are exhibited in Fig.~\ref{fig_res1}, where the first two rows display enlarged local images indicated by color boxes, and the last four rows show the WSIs of samples. The dataset in use includes the original autofluorescence WSIs of different samples and their consecutive registered H$\&$E stained slices (i.e., adjacent layers of the sliced sample). The initial autofluorescence slice is shown in Fig.~\ref{fig_res1}(a), while the consecutive H$\&$E stained slice is shown in Fig.~\ref{fig_res1}(b) as the reference. The H$\&$E virtual staining results using the original CycleGAN \cite{Zhu2017} are represented in Fig.~\ref{fig_res1}(c), showing low conversion accuracy of color (with some areas exhibiting color reversal) and detailed features. As the results shown in Figs.~\ref{fig_res1}(d) and (e) respectively, the method proposed by Wang et al. \cite{Wang2020Virtual} mitigates the color reversal problem and improves  transformation quality to some extent by incorporating a multiscale SSIM loss of green channel into the original CycleGAN network, while the UTOM method \cite{Li2021} achieves superior performance by integrating a threshold and mask strategy into the CycleGAN network. The H$\&$E virtual staining results by our method, displayed in Fig.~\ref{fig_res1}(f), indicate that our model effectively handles this task and achieves performance similar to that of the UTOM method in terms of color conversion accuracy and detailed features preservation.

For large-scale image training and tiling, we randomly crop the training set into $512\times 512$ pixel images from the $4608\times 4608$ pixel unpaired autofluorescence WSIs and H$\&$E-stained WSIs for model training, resulting in a dataset of hundreds of patch images. For testing, the $4608\times 4608$ autofluorescence WSI is divided into 1089 patch images, each with 128-pixel overlaps, and then input into the model for virtual staining, ensuring successful confidence-matrix-based large-scale tiling subsequently. Specifically, the weight matrix $\mathcal{W}$ in use is a $4608\times 4608\times 3$ matrix, with consistent weight values for each RGB channel. According to Eq.~(\ref{weight matrix}), each channel of the weight matrix $\mathcal{W}$ is composed with $36\times 36$ small blocks, where each block consists of $128\times 128$ elements with the same numerical value. After performing virtual staining, we place the $N=1089$ patch images into a zero-padding matrix to obtain a sequence of images, Sequence $YP_{1 \sim N}$. We then perform the Hadamard product between Sequence $YP_{1 \sim N}$ and the weight matrix $\mathcal{W}$ according to Eq.~(\ref{synthesized matrix}), resulting in the output stained WSI.

\begin{table}[htbp]
\centering
\caption{Quantitative evaluation of autofluorescence to H$\&$E virtual staining.}
\begin{tabular}{cccccc}
\hline
    & LPIPS$\downarrow$ & FID$\downarrow$& IS $\uparrow$ & NIQE$\downarrow$& Corr.$\uparrow$ \\
\hline
CycleGAN (baseline) & 0.52 & 235.91 & 2.16& 4.28 & 0.59\\
Wang et al. & 0.48 & 145.41 & 4.59& 4.40 & 0.44\\
UTOM & 0.46 &126.91  & 6.73 & 3.42 &0.68\\
Ours &  \textbf{0.43}  & \textbf{114.19}  &\textbf{7.43} & \textbf{3.27} & \textbf{0.72}\\
\hline
\end{tabular}
  \label{tab:res1}
\caption*{Corr.: Histogram correlation; $\downarrow$: the lower the better; $\uparrow$: the higher the better; the bold font indicates the best performance.}
\vspace{-0.5cm}
\end{table}

Following confidence-matrix-based tiling, our method well eliminates the tiling artifacts, as demonstrated by the WSIs shown in the Fig.~\ref{fig_res1}. In addition to visual perception, we also calculate some quantitative evaluation metrics to compare large-scale virtual staining performance across different methods. Since the reference H$\&$E stained slice in Fig.~\ref{fig_res1}(b) is the adjacent layer rather than the exact ground truth of the input autofluorescence WSI in Fig.~\ref{fig_res1}(a), we use metrics suited for situations without ground truths, such as Learned Perceptual Image Patch Similarity (LPIPS) \cite{Zhang2018lpips}, Fréchet Inception Distance (FID) \cite{Heusel2017}, Inception Score (IS) \cite{Salimans2016}, Naturalness Image Quality Evaluator (NIQE) \cite{Mittal2012}, and histogram correlation\cite{McAndrew2004}. Table~\ref{tab:res1} shows that our method achieves the best results in all quantitative indicators, as marked by the bold font.

\begin{figure}[htb]
\centering
{\includegraphics[width=0.95\linewidth]{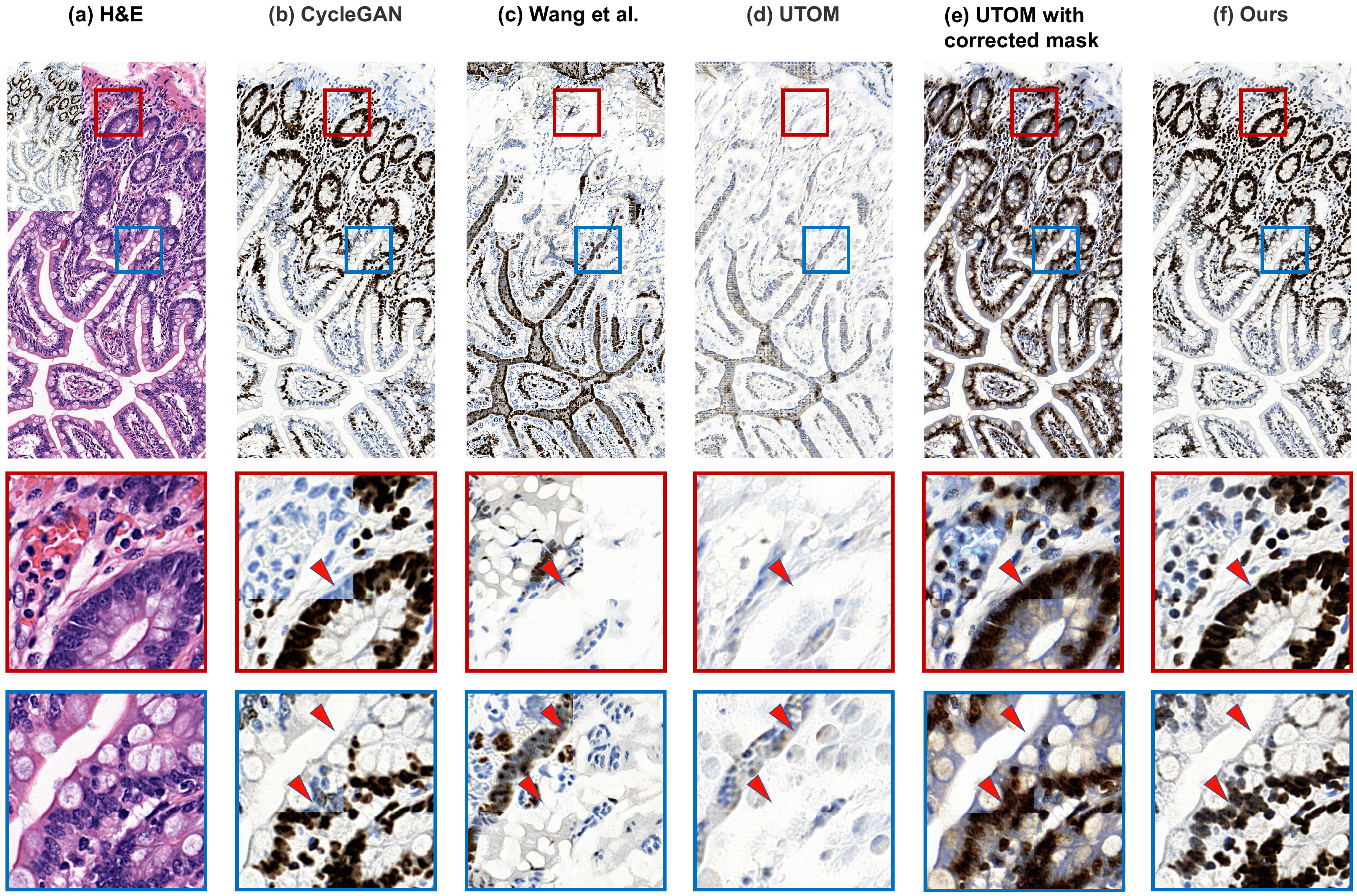}}
\caption{Virtual re-staining from H$\&$E to IHC(Ki67). (a) shows the input real H$\&$E-stained slice, with the real Ki67 stained slice in the top left corner. (b), (c), (d), and (e) display the virtual IHC(Ki67) stained results obtained using the original CycleGAN, the method by Wang et al., the original UTOM, and UTOM with corrected mask, respectively. (f) represents the virtual IHC(Ki67) stained results obtained using our proposed VM-GAN method. Red arrows indicate coloring errors and inconsistent artifacts between patches. Our method effectively addresses the problem of inconsistent staining of similar spatial structures.}
\label{fig_res_ki}
\rule{\textwidth}{1pt}
\end{figure}

\subsection{H$\&$E to IHC(Ki67)}

H$\&$E stained slides typically contain basic morphological information cells, tissues, and tissue blocks \cite{Ghaznavi2013,Wittekind2003}, but they lack microscopic molecular information in cells, such as antigen (protein) expression, which hinders further analysis and evaluation. IHC testing binds target antibodies to corresponding proteins and highlights these protein-bound antibodies with different colored reagents, visualizing specific proteins on tissue slides \cite{Ramos-Vara2014,Xu2019}. Converting H$\&$E stained slices to IHC stained sections by using deep learning techniques is both convenient and cost-efficient for subsequent diagnostic analysis. Therefore, we demonstrate our method by transforming H$\&$E stained slices into IHC(Ki67) stained sections. Ki67 is an important cancer antigen and a clear proliferative marker that can help doctors determine a patient's cancer prognosis or recovery opportunities.

We perform a similar workflow as described above to re-stain the H$\&$E WSIs into Ki67 WSIs, without adjusting any network hyperparameters. We calculate the weight matrix based on the current dataset of WSIs, which has a 7168x3584 pixel size, for confidence-matrix-based tiling. As the results shown in Fig.~\ref{fig_res_ki}, we also compare the performance of our method with some widely-used methods. The initial H$\&$E stained images are shown in Fig.~\ref{fig_res_ki}(a) with the WSI in the first row and enlarged images in the next two rows, while the real Ki67 stained slice is shown in the top left corner. It is important to note that the real Ki67 slice is re-stained from the corresponding H$\&$E stained slice. During the cleaning and chemically re-staining process, some structural deformation inevitably occurs, resulting in the final Ki67 slice not being a strict correspondence with H$\&$E slice but only serving as a reference for a rough color pattern. The virtual staining Ki67 images by different methods are presented in Figs.~\ref{fig_res_ki}(b) to (f), respectively. The CycleGAN method and the method proposed by Wang et al. both introduce incorrect coloring in some areas and show heavy block effects, as indicated by red arrows in Figs.~\ref{fig_res_ki}(b) and (c). 

The result obtained using the UTOM method, as shown in Fig.~\ref{fig_res_ki}(d), exhibits severe distortion or more precisely speaking color reversal in the virtual re-staining. This issue arises because the original mask and threshold in UTOM are designed for converting a single-channel (gray) image to a three-channel (RGB) one. Consequently, the extraction of saliency masks is intended to map the image content area to 1 and the background to 0. Specifically, for the gray-scale image, the saliency masks are extracted directly using the equation $\text{sigmoid}[100(a-\alpha)]$ (i.e., parameterized segmentation operators), where '$a$' is the gray-scale image and '$\alpha$' is the threshold for saliency segmentation. However, in current situation, to maintain the original intent of the UTOM for virtual re-staining, we need to modify the parameterized segmentation operators to $1-\text{sigmoid}[100(a-\alpha)]$ to address the color reversal problem. We refer to our modified UTOM method as 'UTOM with corrected mask', as the results shown in Fig.~\ref{fig_res_ki}(e). Even with the mask correction, incorrect coloring phenomenon persists in the re-staining results. In contrast to the inconsistent staining of similar spatial structures observed with existing methods, our model demonstrates stable performance in this re-staining process, effectively overcoming both the color reversal phenomenon and the block effect, as illustrated in Fig.~\ref{fig_res_ki}(f).

\subsection{IHC(Cc10) to IHC(CD31)}
\begin{figure}[htbp]
\centering
{\includegraphics[width=0.85\linewidth]{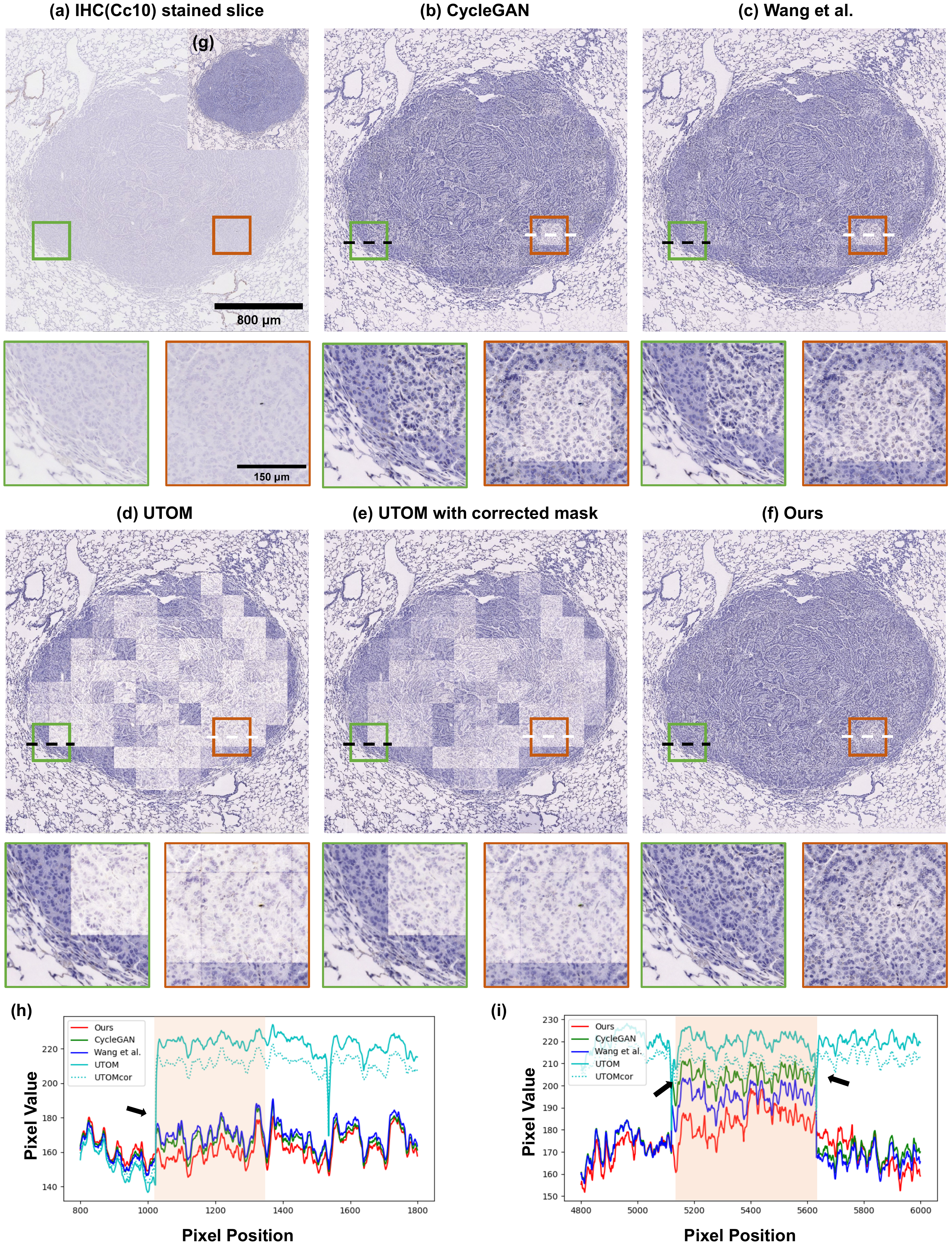}}
\caption{Virtual re-staining from IHC(Cc10) to IHC(CD31) on the ANHIR lung lesion dataset. (a) is the initial Cc10 stained slice. (b) to (f) correspond to the virtual re-staining results obtained using different methods, including the original CycleGAN, the method by Wang et al., the original UTOM, UTOM with corrected mask, and our method. The scale bar represents 800 $\mu$m for the WSIs and 150 $\mu$m for the enlarged images. (g) is the real CD31 stained slice used as the reference. (h) and (i) represent the line profiles of the black and white dashed lines in the corresponding WSIs, respectively. This experiment proves that our method effectively addresses the problem of inconsistent staining contrast between patches and their surrounding. UTOMcor: UTOM with corrected mask.}
\label{fig_res_cc}
\rule{\textwidth}{1pt}
\end{figure}
To further exhibit the generality of our VM-GAN method, we conduct re-staining experiments on different types of IHC data, specifically selecting the Cc10 \cite{Linnoila2000} and CD31 \cite{Liu2012CD31} modalities from the lung lesion subset of the ANHIR dataset. In a larger 7168x7168 pixel area, we similarly train the data after cropping and then adjust the weight matrices based on the input image size for post-processing. The results are shown in Fig.~\ref{fig_res_cc}, where the Fig.~\ref{fig_res_cc}(a) is the initial Cc10 stained slice and Fig.~\ref{fig_res_cc}(g) is the targeted CD31 stained slice. The results from Figs.~\ref{fig_res_cc}(b) to (f) display the virtual stained CD31 images obtained using different methods.

Owning to both the VM-GAN network and the confidence tiling scheme, our method exhibits better spatial consistency in large-scale virtual staining, while the results of other methods show inconsistencies in color, contrast, and brightness between some patches and the surrounding areas. Similarly, the real CD31 slice in Fig.~\ref{fig_res_cc}(g) is chemically re-stained from the corresponding Cc10 slice and inevitably suffers some structural deformation. Thus, the final obtained CD31 slice does not strictly correspond to the Cc10 slice in Fig.~\ref{fig_res_cc}(a) but serves only as a reference.

For further quantitatively evaluation, we select two regions from the re-stained WSI results to plot the line profiles and analyze the pixel intensity changes, as shown in Figs.~\ref{fig_res_cc}(h) and (i). As indicated by the arrows in these figures, our method maintains relatively stable continuity of intensity values, whereas other methods exhibit drastic numerical jumps. In addition to the visual perception in Fig.~\ref{fig_res_cc}, we also calculate quantitative metrics to evident the block-effect elimination and high-quality color transformation performance in the tiled WSIs achieved by our method, as shown in Table~\ref{tab:res2}. According to these metrics, our method shows superior image generation quality and staining realism, outperforming almost all other methods.
\begin{table}[htbp]
\centering
\caption{Quantitative evaluation of IHC(Cc10) to IHC(CD31) virtual staining.}
\begin{tabular}{cccccc}
\hline
  & LPIPS$\downarrow$\cite{Zhang2018lpips} & FID$\downarrow$\cite{Heusel2017} & IS $\uparrow$\cite{Salimans2016} & NIQE$\downarrow$\cite{Mittal2012} & Corr.$\uparrow$ \\
\hline
CycleGAN & 0.43 & 63.55 & 25.61& 3.58 & 0.91\\
Wang et al. & 0.43 & 64.62 & 25.16& 3.71 & 0.97\\
UTOM & 0.43 &55.79  & 23.02 & \textbf{3.03} &0.91\\
UTOM with corrected  & 0.42 &72.89  & 32.67 & 3.34 &0.93\\
Ours &  \textbf{0.41}  & \textbf{36.49}  &\textbf{43.61} & 3.63 & \textbf{0.99}\\
\hline
\end{tabular}
  \label{tab:res2}
\caption*{Corr.: Histogram correlation; $\downarrow$: the lower the better; $\uparrow$: the higher the better; the bold font indicates the best performance.}
\vspace{-0.5cm}
\end{table}

Based on the results of the aforementioned experiments, our method demonstrates robust performance in virtual staining without the need to adjust network hyperparameters. It consistently achieves near-optimal staining performance across various scenarios. In contrast, the existing methods we compare against only exhibit good performance in specific scenarios and often underperform in each scenario relative to our method. Furthermore, by incorporating the proposed confidence tiling scheme, our method achieves outstanding results in inter-modal and large-scale virtual staining.

\section{Conclusion}
In summary, we propose a virtual staining method for microscopic imaging based on the CycleGAN network, enhanced by a newly designed value loss, which we refer to as the VM-GAN network. The proposed value loss ensures accurate color and structure mapping relationships between the input and output images during color transformation. When dealing with large-scale WSI staining, a process of patch splitting, virtual staining, and final tiling is necessary. Existing methods typically face the problem of square effects between patches. To address this, we adopt a plug-and-play post-processing method called the confidence tiling scheme, which effectively improves the continuity of stained patches and eliminates tiling artifacts. We conduct multiple experiments to demonstrate the robustness and efficiency of our method through virtual staining between three different modalities,  such as autofluorescence to H$\&$E, H$\&$E to IHC(Ki67), and IHC(Cc10) to IHC(CD31). Our method exhibits excellent generality and large-scale imaging capability. We believe this work has significant reference value for the future application of microscopic imaging and pathological analysis.

\subsection*{Disclosures}
The authors declare that there are no conflicts of interest related to this article.

\subsection* {Data Availability Statement} 
The data that support the findings of this study are available from the corresponding author upon reasonable request.

\subsection* {Acknowledgments}
This work was supported by the National Natural Science Foundation of China (Grant Nos. 62071219, 62371006), and China Postdoctoral Science Foundation (Grant No. GZC20230057).


\bibliography{report}   

\begin{thebibliography}{10}

\bibitem{Bancroft2008}
J.~D. Bancroft and M.~Gamble, {\em Theory and practice of histological techniques}, Elsevier health sciences  (2008).

\bibitem{Titford2005}
M.~Titford, ``The long history of hematoxylin,'' {\em Biotechnic \& Histochemistry} {\bf 80}(2), 73--78  (2005).

\bibitem{Coudray2018}
N.~Coudray, P.~S. Ocampo, T.~Sakellaropoulos, {\em et~al.}, ``Classification and mutation prediction from non--small cell lung cancer histopathology images using deep learning,'' {\em Nature Medicine} {\bf 24}(10), 1559--1567  (2018).

\bibitem{Ramos-Vara2014}
J.~Ramos-Vara and M.~Miller, ``When tissue antigens and antibodies get along: revisiting the technical aspects of immunohistochemistry—the red, brown, and blue technique,'' {\em Veterinary Pathology} {\bf 51}(1), 42--87  (2014).

\bibitem{Barbastathis2019}
G.~Barbastathis, A.~Ozcan, and G.~Situ, ``On the use of deep learning for computational imaging,'' {\em Optica} {\bf 6}(8), 921--943  (2019).

\bibitem{Weigert2018}
M.~Weigert, U.~Schmidt, T.~Boothe, {\em et~al.}, ``Content-aware image restoration: pushing the limits of fluorescence microscopy,'' {\em Nature Methods} {\bf 15}(12), 1090--1097  (2018).

\bibitem{Wang2019}
H.~Wang, Y.~Rivenson, Y.~Jin, {\em et~al.}, ``Deep learning enables cross-modality super-resolution in fluorescence microscopy,'' {\em Nature Methods} {\bf 16}(1), 103--110  (2019).

\bibitem{Ouyang2018}
W.~Ouyang, A.~Aristov, M.~Lelek, {\em et~al.}, ``Deep learning massively accelerates super-resolution localization microscopy,'' {\em Nature Biotechnology} {\bf 36}(5), 460--468  (2018).

\bibitem{Wu2019}
Y.~Wu, Y.~Luo, G.~Chaudhari, {\em et~al.}, ``Bright-field holography: cross-modality deep learning enables snapshot 3d imaging with bright-field contrast using a single hologram,'' {\em Light: Science \& Applications} {\bf 8}(1), 25  (2019).

\bibitem{Ounkomol2018}
C.~Ounkomol, S.~Seshamani, M.~M. Maleckar, {\em et~al.}, ``Label-free prediction of three-dimensional fluorescence images from transmitted-light microscopy,'' {\em Nature Methods} {\bf 15}(11), 917--920  (2018).

\bibitem{Christiansen2018}
E.~M. Christiansen, S.~J. Yang, D.~M. Ando, {\em et~al.}, ``In silico labeling: predicting fluorescent labels in unlabeled images,'' {\em Cell} {\bf 173}(3), 792--803  (2018).

\bibitem{Zhu2017}
J.-Y. Zhu, T.~Park, P.~Isola, {\em et~al.}, ``Unpaired image-to-image translation using cycle-consistent adversarial networks,'' in {\em Proceedings of the IEEE International Conference on Computer Vision},  2223--2232  (2017).

\bibitem{Bai2023}
B.~Bai, X.~Yang, Y.~Li, {\em et~al.}, ``Deep learning-enabled virtual histological staining of biological samples,'' {\em Light: Science \& Applications} {\bf 12}(1), 57  (2023).

\bibitem{Boyd2022}
J.~Boyd, I.~Villa, M.-C. Mathieu, {\em et~al.}, ``Region-guided cyclegans for stain transfer in whole slide images,'' in {\em International Conference on Medical Image Computing and Computer-Assisted Intervention},  356--365, Springer  (2022).

\bibitem{Liu2021}
S.~Liu, B.~Zhang, Y.~Liu, {\em et~al.}, ``Unpaired stain transfer using pathology-consistent constrained generative adversarial networks,'' {\em IEEE Transactions on Medical Imaging} {\bf 40}(8), 1977--1989  (2021).

\bibitem{Wang2020Virtual}
R.~Wang, P.~Song, S.~Jiang, {\em et~al.}, ``Virtual brightfield and fluorescence staining for fourier ptychography via unsupervised deep learning,'' {\em Optics Letters} {\bf 45}(19), 5405--5408  (2020).

\bibitem{Li2021}
X.~Li, G.~Zhang, H.~Qiao, {\em et~al.}, ``Unsupervised content-preserving transformation for optical microscopy,'' {\em Light: Science \& Applications} {\bf 10}(1), 44  (2021).

\bibitem{Musumeci2014}
G.~Musumeci, ``Past, present and future: overview on histology and histopathology,'' {\em J Histol Histopathol} {\bf 1}(5), 1--3  (2014).

\bibitem{Lee2023}
C.~Lee, G.~Song, H.~Kim, {\em et~al.}, ``Deep learning based on parameterized physical forward model for adaptive holographic imaging with unpaired data,'' {\em Nature Machine Intelligence} {\bf 5}(1), 35--45  (2023).

\bibitem{Liu2024}
Z.~Liu, L.~Chen, H.~Cheng, {\em et~al.}, ``Virtual formalin-fixed and paraffin-embedded staining of fresh brain tissue via stimulated raman cyclegan model,'' {\em Science Advances} {\bf 10}(13), eadn3426  (2024).

\bibitem{Stockman2001}
G.~Stockman and L.~G. Shapiro, {\em Computer vision}, Prentice Hall PTR  (2001).

\bibitem{Bargshady2020}
G.~Bargshady, X.~Zhou, R.~C. Deo, {\em et~al.}, ``The modeling of human facial pain intensity based on temporal convolutional networks trained with video frames in hsv color space,'' {\em Applied Soft Computing} {\bf 97}, 106805  (2020).

\bibitem{Farahani2015}
N.~Farahani, A.~V. Parwani, and L.~Pantanowitz, ``Whole slide imaging in pathology: advantages, limitations, and emerging perspectives,'' {\em Pathology and Laboratory Medicine International} , 23--33  (2015).

\bibitem{Liu2020}
Y.~Liu, X.~Li, A.~Zheng, {\em et~al.}, ``Predict ki-67 positive cells in h\&e-stained images using deep learning independently from ihc-stained images,'' {\em Frontiers in Molecular Biosciences} {\bf 7}, 183  (2020).

\bibitem{Borovec2020}
J.~Borovec, J.~Kybic, I.~Arganda-Carreras, {\em et~al.}, ``Anhir: automatic non-rigid histological image registration challenge,'' {\em IEEE Transactions on Medical Imaging} {\bf 39}(10), 3042--3052  (2020).

\bibitem{Lu2016}
F.-K. Lu, D.~Calligaris, O.~I. Olubiyi, {\em et~al.}, ``Label-free neurosurgical pathology with stimulated raman imaging,'' {\em Cancer Research} {\bf 76}(12), 3451--3462  (2016).

\bibitem{Orringer2017}
D.~A. Orringer, B.~Pandian, Y.~S. Niknafs, {\em et~al.}, ``Rapid intraoperative histology of unprocessed surgical specimens via fibre-laser-based stimulated raman scattering microscopy,'' {\em Nature Biomedical Engineering} {\bf 1}(2), 0027  (2017).

\bibitem{Hollon2020}
T.~C. Hollon, B.~Pandian, A.~R. Adapa, {\em et~al.}, ``Near real-time intraoperative brain tumor diagnosis using stimulated raman histology and deep neural networks,'' {\em Nature Medicine} {\bf 26}(1), 52--58  (2020).

\bibitem{Zhang2018lpips}
R.~Zhang, P.~Isola, A.~A. Efros, {\em et~al.}, ``The unreasonable effectiveness of deep features as a perceptual metric,'' in {\em Proceedings of the IEEE Conference on Computer Vision and Pattern Recognition},  586--595  (2018).

\bibitem{Heusel2017}
M.~Heusel, H.~Ramsauer, T.~Unterthiner, {\em et~al.}, ``Gans trained by a two time-scale update rule converge to a local nash equilibrium,'' {\em Advances in Neural Information Processing Systems} {\bf 30}  (2017).

\bibitem{Salimans2016}
T.~Salimans, I.~Goodfellow, W.~Zaremba, {\em et~al.}, ``Improved techniques for training gans,'' {\em Advances in Neural Information Processing Systems} {\bf 29}  (2016).

\bibitem{Mittal2012}
A.~Mittal, R.~Soundararajan, and A.~C. Bovik, ``Making a “completely blind” image quality analyzer,'' {\em IEEE Signal Processing Letters} {\bf 20}(3), 209--212  (2012).

\bibitem{McAndrew2004}
A.~McAndrew, {\em An introduction to digital image processing with MATLAB}, Course Technology Press  (2004).

\bibitem{Ghaznavi2013}
F.~Ghaznavi, A.~Evans, A.~Madabhushi, {\em et~al.}, ``Digital imaging in pathology: whole-slide imaging and beyond,'' {\em Annual Review of Pathology: Mechanisms of Disease} {\bf 8}, 331--359  (2013).

\bibitem{Wittekind2003}
D.~Wittekind, ``Traditional staining for routine diagnostic pathology including the role of tannic acid. 1. value and limitations of the hematoxylin-eosin stain,'' {\em Biotechnic \& Histochemistry} {\bf 78}(5), 261--270  (2003).

\bibitem{Xu2019}
Z.~Xu, X.~Huang, C.~F. Moro, {\em et~al.}, ``Gan-based virtual re-staining: a promising solution for whole slide image analysis,'' {\em arXiv preprint arXiv:1901.04059}   (2019).

\bibitem{Linnoila2000}
R.~I. Linnoila, E.~Szabo, F.~DeMayo, {\em et~al.}, ``The role of cc10 in pulmonary carcinogenesis: from a marker to tumor suppression,'' {\em Annals of the New York Academy of Sciences} {\bf 923}(1), 249--267  (2000).

\bibitem{Liu2012CD31}
L.~Liu and G.-P. Shi, ``Cd31: beyond a marker for endothelial cells,'' {\em Cardiovasc Res} {\bf 94}(1), 3--5  (2012).

\end{thebibliography}
\bibliographystyle{spiejour}   


\vspace{2ex}\noindent\textbf{Junjia Wang} is a graduate student in the School of Electronic Science and Engineering, Nanjing University. He received his B.S. degree from Northwestern Polytechnical University in 2021. His research interests include computational photography and virtual staining.

\vspace{1ex}
\noindent Biographies and photographs of the other authors are not available.

\listoffigures
\listoftables

\end{spacing}
\end{document}